\documentstyle[11pt,moriond,epsfig]{article}

\def\Journal#1#2#3#4{{#1} {\bf #2}, #3 (#4)}

\def\ASR{\em Adv. Space Res.}
\def\RSI{\em Rev. Sci. Instr.}
\def\CRY{\em Cryogenics}
\def\AO{\em Appl. Optics}
\def\APJ{\em Astrophys. J.}
\def\IRP{\em Infrared Phys.}

\def\PRD{{\em Phys. Rev.} D}

\def\be{\begin{equation}}
\def\ee{\end{equation}}
\def\bea{\begin{eqnarray}}
\def\eea{\end{eqnarray}}

\begin{document}
\vspace*{4cm}
\title{CMB ANISOTROPY MEASUREMENTS USING STRATOSPHERIC BALLOONS}

\author{ P. DE BERNARDIS, S. MASI}

\address{Department of Physics, University of Rome La Sapienza, \\
P.le A. Moro 2, 00185 Roma, Italy}

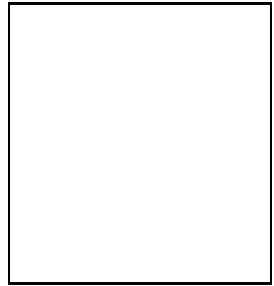
\maketitle\abstracts{ We review the topic of Cosmic Microwave Background
Anisotropy measurements carried out by means of balloon-borne telescopes.
After a short description of the experimental methodology, we outline 
the peculiar problems of these experiments, and we describe the
main results obtained and the perspects for future developments.
}

\section{Stratospheric Balloons and CMB Anisotropy Measurements}

\subsection{Stratospheric Balloons}\label{subsec:sb}

Stratospheric balloons are space carriers offering 
flights at 30-40 Km altitude, for payloads with mass
up to 2500 Kg, with reasonable cost and short development
time scales. The available time at float ranges between a few hours and
a couple of weeks. The intrinsic payload stability is
of the order of a few arcminutes. The observable elevations
are between 0$^o$ and 65$^o$ (or up to the zenith placing the
payload on top of the balloon). 

The team of NASA-NSBF (USA) offers 6-48 hours flights over 
North America and 1-2 weeks flights in Artic and Antarctic 
(Long Duration Balloons). The team of 
CNES (France) offers balloon flights of a few hours over
France and up to 60 hours in Sweden. The team of ASI (Italy)
offers 20-40 hours transmediterranean flights from Sicily
to Spain. Balloon flights are also carried out by teams
in Russia, Brazil, Japan and in China.

Infrared mongolfieres\cite{Ma} have been used for ultra long duration
flights of lightweight payloads ($\sim$ 50 Kg).
Superpressure balloons are being developed, providing in the near
future ultra-long  ($\sim $ 3 months) balloon flights for medium-mass 
payloads. 

\subsection{CMB measurements from Stratospheric 
Balloons}\label{subsec:CMBS}

The main advantage of placing a CMB telescope on a stratospheric 
balloon is a large reduction of the atmospheric brightness (temperature) 
and noise. This effect is wavelength dependent. The advantage is huge for 
observation wavelengths above 100 GHz, while it is not too important
for wavelenghts below 40 GHz, where longer observation
time and easier access to the experiment favour high mountain
observations. For wide band bolometric receivers the balloon
environment offers an additional advantage, 
since the reduced radiation background
permits to reach a lower temperature of the sensor, 
thus increasing significantly its sensitivity.

A generic instrument for the measurement of CMB anisotropies
is composed by the following main subsystems:
the gondola, usually an aluminum or carbon fiber tubolar 
lightweight structure with an inner frame (containing the telescope,
the receiver and a star sensor for attitude reconstruction) 
and an outer frame containing the elevation mechanism 
for the inner frame, the azimuth mechanism and a pivot to 
rotate the entire payload with respect to the flight chain, 
the data processing and storage electronics and 
lithium batteries to power the system.
The space agency provides the uplink and downlink telemetry 
and the navigation hardware (GPS, altitude sensors, ballast, 
balloon valves etc..).

Placing the telescope below a stratospheric balloon has a number
of disadvantages. Interactivity with the experiment is greatly
reduced, and the duration of the measurement is only
a few hours for normal balloon flights. So all the tests for systematic
effects must be planned in advance, with great care, and inserted
in the flight sequence, taking advantage of the increased sensitivity 
of balloon systems to perform quicker test. Also, the balance between
time devoted to integration, time for calibration, and tests 
for systematics must be optimized. Long Duration Ballooning
offers unique extensive possibilities of systematics tests.

The telescope must point/scan the sky regions selected for
calibration and observations. Pointing a telescope attached to
a stratospheric balloon is much more difficult than pointing
a ground based telescope, and also more difficult than 
pointing a satellite, due to the presence of gravity
and of perturbations induced by the residual atmosphere.
A huge effort has been paid by X-ray and IR astronomers
to develop Attitude Control Systems (ACS) suitable for balloon
borne experiments\cite{Ho} \cite{Ch} \cite{Fx} \cite{Lam} \cite{Fi}. 
These systems are based on
gyroscopes, magnetometers and star cameras as sensors,
and on motorized high inertia flywheels as actuators
to control the payload motion. 
Receiver calibration is usually achieved observing a planet.
The performance of the ACS should be good enough to provide
a pointing stability of at least FWHM/20, with the additional
necessity of making good raster scans for beam shape
and calibration purposes. 
An advantage of balloon experiments with respect to 
ground based experiments is the possibility of
measuring the large scale dipole anisotropy of the CMB,
which provides a calibrator with the same spectrum of
the (sub-)degree scale anisotropies. In this case,
the ACS should be able to rotate the experiment over
360$^o$ in azimuth, with a scan speed large enough to
mantain the dipole signal inside the useful electrical 
bandwidth of the instrument and with negligible pendulations.
Due to atmospheric inhomogeneity and sidelobes spillover,
this calibration mode is practically impossible for 
ground-based experiments.

The balloon - flight chain - payload system is a pendulum.
The balloon drifts with the stratospheric jet-stream.
Inhomogeneities in the wind can excite pendulations of the
system, which has three main pendulation modes:
a rigid pendulation of the system (period $\sim 30 s$); 
a pendulation of the payload in opposition with the balloon
(period $\sim 10 s$), and a pendulation of the gondola
around the pivoting point (period o a few s).
Pendulations of the experiment can also be induced
by the attitude control system itself, due to the
non-diagonal nature of the inertia tensor of the payload
and to static and dynamic unbalancing of the flywheels. 
Pendulations modulate the residual atmospheric brightness,
thus producing a periodic signal in the receiver.
If $I_{ATM}(e, \nu)$ is the atmospheric brightness, $E(\nu)$ is
the spectral efficiency, and $e$ is the beam elevation,
pendulations $\Delta e$ produce brightness fluctuations equivalent to
CMB temperature fluctuations, of the order of
$$
{\Delta T_{CMB} \over \Delta e} =
{T_{CMB} \over \tan e}
{ \int I_{ATM}(e, \nu) E(\nu) d\nu
\over 
\int B (T_{CMB}, \nu) {xe^x \over e^x -1} E(\nu) d\nu}
\eqno.
$$
Bolometric systems are prone to this problem, since
their spectral efficiency is intrinsically quite wide,
and the $E(\nu)$ is defined by means of suitable
blocking / mesh filters. 
In fig.1 we compare the spectrum of the atmosphere
to the spectrum of the CMB anisotropy as seen through
a suitable filters chain defining a 90 GHz nominal 
bandpass. The pendulation induced signal can be reduced
to $\sim 1 \mu K / arcmin$ with extreme care in the
quality and choiche of the receiver filters, and with
good performance of the ACS (pendulations lower than
a few arcmin peak to peak).

Radio Frequency interference due to powerful telemetry
transmitters is difficult to fight, especially for 
the ultrasensitive bolometric microwave receivers
used for CMB anisotropy research. To understand the
extent of the problem, one should notice that the
telemetry transmitters have a power of a few W at
a frequency of a few GHz, while the CMB receivers
have a sensitivity of the order of 10$^{-17}$ W
at a frequency only 100 times higher. Extensive
filtering of the signal lines exiting the receiver
and wavelength selection / filtering on the optical path
of the receiver must be carefully optimized.

The problem of sidelobes rejection is common to all
the CMB experiments. These experiments try to detect
$~10 \mu K$ signals embedded in a 300K environment.
Balloon experiments cannot take advantage of drift scans,
due to the very long integration time required by
such an observation mode.
Very clean (low sidelobes) off-axis telescopes and shielding 
systems have been developed \cite{Fi} \cite{Ge} \cite{TopHat} . 
The problem is more severe for
long wavelength HEMT-based recievers.
Measurements of the same sky patches carried out at
different locations during the balloon flight are the only 
way to test for sidelobes contamination in the cosmological
data. In this respect balloon experiments cannot compete
with the planned lagrangian point space missions \cite{Be} \cite{map}.
Anyway, no evidence for spillover contamination has been
found in sensitive balloon-borne CMB surveys.

Balloon borne experiments devoted to CMB anisotropy 
(recent, ongoing and planned) are listed in table 1.
The general trend is towards higher angular resolution. Sub-degree
resolutions allow to detect the acoustic peaks in the power spectrum 
of CMB anisotropy, inferring cosmological parameters,
and testing for gaussian versus non gaussian CMB fluctuations.
The detectability of higher $\ell$ acoustic peaks depends on
both the angular resolution and noise performance of the instrument.
In general, a beam FWHM smaller than 20 arcmin is required 
to start to make a reasonable $\ell$-space spectroscopy of the acoustic 
peaks. 10 arcmin FWHM beams are optimal.

First generation experiments made statistical samples of the sky temperature
fluctuations $\Delta T_i$. Second generation experiments are making maps of 
the CMB temperature field $\Delta T (RA,dec)$, thus producing a lot more information.
Transition from the first to the second generation of experiments
has been possible because of the development of new detectors: 
Spider-web bolometers and sensitive HEMTs receivers. 
The two techniques are in competition. Also, the map making
experiments are based on scanning techniques rather than on
pointed observations. 

\subsection{Bolometers versus HEMTs}\label{subsec:bolhemts}

High Electron Mobility Transistors (HEMTs) are used to make sensitive, 
wide band (10 GHz), coherent detectors\cite{Pop}. They work up to 50 GHz, with good 
noise performances: $NET_{CMB} \sim 500~\mu K \sqrt{s}$ is the state 
of the art for the noise figure. 
These devices work at reasonable cryogenic temperatures (4 K) and feature 
high speed and high rejection of out of band radiation. As a result, they 
are quite insensitive to atmospheric emission and fluctuations. 
Also, the fast response allows for fast scans of the sky, with 
tendency towards large sky coverage. Working at 
relatively low frequencies, they need large telescopes to get a given 
beamsize: getting a diffraction limited beam of 10 arcmin FWHM at
40 GHz requires a 3 m diameter telescope. 
At higher frequencies (up to 150 GHz), 
coherent detection systems use SIS junctions
as mixers between CMB radiation and radiation from a stable local oscillator.
The down-converted intermediate frequency signal is amplified by 
low noise cryogenic HEMTs and detected. 

Cryogenic Bolometers are used to make incoherent, wide band (30-100 GHz) 
detectors, operating at frequencies higher than 90 GHz with extremely low 
noise : $NET_{CMB} \sim 100~\mu K \sqrt{s}$ 
is the state of the art. They need very low temperatures 
(0.3K-0.1K), so require complex cryogenic systems, which must
be expecially rugged in the case of balloon-borne experiments 
\cite{Che} \cite{Pal} \cite{Ben} \cite{Mas}. 
Being thermal detectors, feature slow response 
(tens of ms) and are sensitive to the load from the radiative background.
The telescope emissivity needs to be really low, and spillover from
ambient temperature blackbody radiation can significantly degrade the
bolometer performances. Moreover, bolometers need sophisticated out of band  
radiation rejection / filtering techniques \cite{Page} \cite{Ade}. 
Working at higher frequencies, they feature higher angular resolution 
for the same telescope size: a 0.9 m diameter telescope at 150 GHz 
already produces a diffraction limited beam of 10 arcmin FWHM. 
Bolometers are sensitive to any kind of energy deposited in the
sensing element. Ionization by cosmic rays (mainly protons) in the stratosphere
represents an annoying source of excess noise in bolometric receivers.
The problem has been only recently solved with the development of web-like
absorbers\cite{Mau}, which feature negligible cross section for 
cosmic rays while mantaining full sensitivity to CMB photons and 
reduced heat capacity, resulting in quicker response.

\subsection{Sky scan techniques}\label{subsec:skyscan}

The first generation of experiments performed chopped measurements.
The beam direction was moved periodically in the sky, thus allowing 
synchronous detection. The signal to noise ratio was improved spending a long 
integration time for each selected observed direction.
The measured quantities are 
$ \Delta T_{meas,i} = {1 \over \tau} \int_0^\tau \Delta T ( {\vec x}(t) ) R(t) dt$
and the mean square value of the measured signals was compared to the
theoretical $<\Delta T^2 > = \sum_\ell {2 \ell + 1 \over 4 \pi} c_\ell w_\ell$.
Here the window function $w_\ell$ depends on the beam shape and on the
details of the modulation performed by the
chopper and of the demodulation performed by the signal processing electronics
in the receiver. For example, for a total power gaussian beam experiment 
$w_\ell = \exp [ -{1\over 2} \ell (\ell+1) \sigma_B^2 ]$,
while for a square wave chop with peak to peak amplitude $\alpha$,
$w_\ell = 2 [ 1-P_\ell(\cos \alpha ) ] \exp [ -{1\over 2} \ell (\ell+1) \sigma_B^2 ]$.
The results of these experiments are one or a few (if multiple window functions
can be obtained from the signals processing) data points in the CMB anisotropy
power spectrum. Typical combined analysis of results from these experiments can
be found e.g. in  \cite{Bal} \cite{Web} \cite{Rat}.

The second, new generation of experiments performs scan measurements.
The telescope beam scans the sky at constant speed. Different spherical harmonic 
components of the CMB temperature field produce different 
electrical frequencies in the detector. This experimental approach requires 
extremely low detector noise and extremely high system stability.
$\ell$-space spectroscopy 
is possible using a scanning instrument and an averaging signal analyzer. 
The procedure to get power spectra from these data is outlined in \cite{Del}.
Sky maps can also be recovered from the time ordered scan data.  
However, drifts and $1/f$ noise
project in the map in a non-trivial way, producing the characteristic
striping well known for the IRAS and DIRBE maps. The best way to
reduce this effect in the data analysis is a hot important topic \cite{Wri0} 
\cite{Jans} \cite{Wri} \cite{Smo} \cite{Teg0} \cite{Teg}, since
the same problem will be present in the data from the future satellite
missions MAP and Planck. It must be stressed that scanning balloon
experiments represent the only available test-bed for 
studying and reducing the effect, experimenting on both the hardware and the software. 
From the experimentalist point of view, the 
first problem is to setup the scan so that the CMB information
is encoded in a frequency band matching the electrical band of
the detection system, and as far as possible from system noise and disturbances. 
Assuming a scan at constant zenith angle $\Theta$, the 
temperature fluctuations of the CMB along the scan can be
expressed as a Fourier series $T(\Theta, \phi) = \sum_m \alpha_m e^{im\phi}$.
If the azimuthal scan rate is $\dot \phi$, 
the $m$-th component of the CMB signal 
(which has a mean square amplitude $\Gamma_m = < \alpha_m \alpha_m^*>$)
will produce in the detector a signal at the electrical frequency 
$f = {\dot \phi} m / 2 \pi$. So a frequency analysis of the detector
signal can be performed to get $m$-space spectroscopy of the CMB
anisotropy. Moreover, the 1-D $m$-space spectrum $\Gamma_m$ is related
in a simple way\cite{Del} to the 2-D $\ell$-space power spectrum $c_\ell$ of the CMB:
$\Gamma_m = \sum_{\ell = |m|}^\infty c_\ell B_\ell^2 P_{\ell m}^2 (\Theta) $.
This relationship is valid for scans along full circles; shorter
scans on circle sections have a lower $\ell$-space resolution.
All the spherical harmonics components
of the CMB with $\ell > m$ contribute to the detector signal
at frequency $m {\dot \phi} / 2 \pi$. If $\ell_{min}$
is the lowest spherical harmonic of interest, the experimentalist will
setup the scan in such a way that noise in the system is confined at
frequencies lower than $f_{min} = \ell_{min} {\dot \phi} / 2 \pi$.
This condition is stricter for the HEMT based receivers, which feature
higher 1/f noise knee. On the other hand, if the highest spherical 
harmonics of interest is $\ell_{max}$ (which depends on the beam size of
the experiment), the scan speed should be adjusted so that the frequency
$f_{max} = \ell_{max} {\dot \phi} / 2 \pi$ is lower than the 
high frequency cut-off of the experiment. This condition
is quite strict for bolometric receivers, which feature $\sim~10~ms$
thermal time constants. HEMT based experiments can afford
faster scan rates, but good tradeoffs are also possible
with bolometer based systems, which, in addition, feature
a lower general noise level. The situation is depicted in fig.2,
where we plot the typical noise spectrum for a balloon borne
CMB experiment, resulting from 1/f and white noise from the 
detector and the atmosphere and broadened lines from pendulation-induced
atmospheric noise. We also plot in fig.2 the transfer function of
the bolometer and the frequency spectrum 
of the CMB signal, as encoded by the scanning procedure and
by the beam of the instrument.
For beam sizes of the order of 20 arcmin, scan speeds of the order 
of 3 deg/s, and bolometer time constants below 100 ms, all the interesting
CMB anisotropy information can be encoded in the lowest noise 
frequency band of the system.

\section{A small zoo of balloon-borne CMB anisotropy experiments}

In the following we summarize the highlights of several recent, ongoing
and planned Balloon measurements of the CMB anisotropy at sub-degree scales 
(see table 1 for a synoptic and for references). It is evident that
widely different approaches are used, mainly because different laboratories
developed and optimized in their past different detection and processing techniques,
and also because we do not know too well the observable to be measured
and the related systematic effects.

\noindent $\bullet$ Flown experiments:

ARGO is a first-generation experiment, 
developed at the University of Rome La Sapienza, ENEA Frascati and IROE-CNR Firenze, 
and flown in 1993. The experiment features a 1.2m cassegrain telescope with 
wobbling secondary, resulting in a 0.8$^o$ FWHM beam chopped by 1.4$^o$ p-p. 
The large throughput bolometric receiver had four detectors with bands centered
at 2.0, 1.2, 0.8, 0.5 mm. CMB anisotropy was detected in the 2.0 and 1.2 mm 
bands in the regions of Hercules and Aries/Taurus, with S/N of the order of
one per pixel (on $\sim$ 200 independent pixels). Spectral ratios and
comparisons with local contaminants templates were essential to asses 
the cosmic nature of the detected fluctuations, excluding local contaminations.

BAM is a first generation experiment
developed by the British Columbia group and flown in 1995. It features a
cryogenic differential Fourier Transform Spectrometer 
with two bolometric receivers, in the
focus of an off axis 1.65 m telescope, resulting in two beams 42 arcmin FWHM 
on the sky separated by 3.6$^o$. Analysis of the rapid-scan interferograms
provides the useful CMB data in five spectral bands (from 93 to 276 GHz), 
plus important checks for systematics. A problem in a memory chip 
reduced the time available for data taking during the flight. CMB anisotropy
was detected in a set of 10 sky directions. A new flight of the payload
is scheduled for May 1998.

BOOMERanG is a second generation experiment, developed by a 
collaboration headed by University of Rome La Sapienza and Caltech,
with contributions from ENEA, IROE-CNR, QMWC, UCSB, U.Mass.
It was flown two times in August 1997 with a total power bolometric 
receiver and a 1.3 m off-axis telescope, 
producing four 20 arcmin FWHM pixels at 150 GHz and two 30 arcmin 
FWHM pixels at 90 GHz. The flight demonstrated the outstanding
performances of system, mainly the low background on the bolometers,
the extremely low level of the sidelobes, and the absence of excess
noise in the total power bolometers readout.
The system performed 40$^o$ wide azimuth scans (forward-back with
a rounded triangle waveform), 
covering a 5$^o$ $\times$ $80^o$ region in Cetus, Aquarius, Capricornus. 
The system was calibrated in flight observing Jupiter and the CMB Dipole.
Custom spider-web bolometers reduced drastically the rate of cosmic ray
events (down to about 60 events in the full flight). 
Excess noise is limited to very low frequencies, unimportant for CMB
anisotropy signals. The data analysis is currently in progress. 
This flight represents also a qualification of the payload for
the long duration flight to be done at the end of this year
(see below).

HACME is a second generation HEMT experiment developed 
by the group of University of California at Santa Barbara. It features
39, 41 and 43 GHz receivers with $\sim 500 \mu K \sqrt{s}$ noise. The gregorian
telescope produces 46 arcmin FWHM beams. 
The most interesting feature of the experiment is a 
rotating flat mirror mounted in front of the telescope. 
The flat is tilted 2.5 degrees relative to its spin axis,
and spins at 2.5 Hz. So the beam scans
the sky on 5$^o$ diameter circles, thus pushing toward large sky coverage,
with very well cross-linked scans. 
The beam elevation changes along the
circle, but the low frequency makes the resulting atmospheric signal negligible.
Also, any residual atmospheric signal at the spin frequency is well outside
the signal frequency band occupied by the degree/scale CMB anisotropy.  
It was flown two times in 1996, observing
about 5000 independent pixels. The disadvantage of this scan technique
is the very short integration time per pixel. Maps recovered from the time
ordered data are already published \cite{Teg}, but the sensitivity
is not high enough to pick-up by-eye cosmic structures from the maps.
The statistical analysis of the data is in progress.

MAX/ACME is a first generation experiment, developed by a 
collaboration between the University of California at Berkeley and
the University of California at Santa Barbara. 
It has been flown five times since 1989. It features a 1.3 m Gregorian 
telescope with wobbling secondary mirror and several multiband bolometric
receivers, resulting in $\sim 0.5^o$ FWHM beamsize at wavelengths
between 3.3 and 0.8 mm. MAX detected CMB anisotropies in 7 different
sky regions, with $\Delta T / T$ ranging from $(1.2^{+0.4}_{-0.3}) \times 10^{-5}$
to $(2.9^{+4.3}_{-1.8}) \times 10^{-5}$. Spectral arguments, taking
advantage of the multiband nature of the system, allow to reject 
the hypothesis of Galactic contamination of the data. All these detections may be 
consistent with coming from a single parent population.

MSAM is a first generation experiment, flown in 1992, 1994, 1995, 
developed by a collaboration
headed by University of Chicago and NASA/Goddard SFC. It features a 
1.4 m off-axis Gregorian telescope with a nutating secondary.
The cryogenic bolometric receiver has four wavelength bands
between 2.0 and 0.42 mm. The instrument had a resolution of 
$\sim 0.5^o$ and sampled a ring centered on the NCP.
Data from the different wavelenght bands were combined
to create separate CMB and dust channels, thus rejecting
local contamination of the cosmological data. 
Repeated measurements of the same sky regions in different flights reproduced
the same anisotropy detection of the CMB, which is also
consistent with the measurements of the ground-based
Saskatoon experiment on the same NCP ring. The second phase
of the MSAM experiment, flown in 1997, used an adiabatic demagnetization
refrigerator to cool five detectors at 0.1K.

QMAP is a HEMT-based second generation experiment developed at Princeton
and University of Pennsylvania. It is basically the Saskatoon 
experiment~\cite{Net} mounted on a balloon platform: 
a chopping flat is mounted in front of an off-axis
1 m primary, feeding two Q-band (39 arcmin FWHM, 8 GHz bandwidth), 
one K-band (54 arcmin FWHM, 8 GHz bandwidth) and one D-band 
(20 arcmin FWHM) receivers. The elevation of the beam is 41$^o$, 
and the chop sweep at 4.4 Hz was 10$^o$ or 20$^o$ depending on the 
flight. The azimuth scan of the gondola produced a second level of 
chop with 10$^o$ amplitude, centered on the NCP. 
The system has been flown in 1996 and 1997, 
and observed about 700 independent pixels in the NCP region.
The data analysis is in progress.

\noindent $\bullet$ Experiments to be flown:

BEAST is a scaled-up version of HACME, developed by UCSB with
contributions from JPL, CNR-Bologna, CNR-Milano, EFEI-INPE (Brazil).
The instrument has a 2.2 m lightweigth primary, with 7 Q-band detectors
(275$\mu K \sqrt{s}$ noise, 19 arcmin FWHM beam) and 2 Ka-band
detectors (180$\mu K \sqrt{s}$ noise, 26 arcmin FWHM beam). 
Using a spinning flat in front of the primary, all
pixels scan on 10$^o$ circles at 5 Hz, while the telescope is
swept through different azimuth angles depending on the mission.
The LDB version of BEAST will cover about 4000 square degrees,
with final sensitivity of about 70 $\mu K$ per 10 arcminutes pixel.
A second LDB flight will have 10 additional W-band (90 GHz) receivers
(500$\mu K \sqrt{s}$ noise, 9 arcmin FWHM beam). The test flight
will be in 1998 and the LDB flights are planned for 1998 and 1999.
ACE is the lightweight Ultra Long Duration version of BEAST, 
planned for 100 days flights.

BOOMERanG LDB is the long duration flight of BOOMERanG N.A.,
currently planned for the Antarctic flight in 1998. The LDB
hardware, including the long duration $^3He$ cryostat, the spider web
bolometers, the ACS sensors and actuators, the sun shield system 
and the fully automated flight programmer / data logger, 
has been tested successfully in the August 1997 flight. The experiment will  
feature an improved focal plane, with 16 detectors: four 90 GHz bolometers
(75$\mu K \sqrt{s}$ noise, 20 arcmin FWHM beam) and
four three-band photometers (90$\mu K \sqrt{s}$ @ 150 GHz, 
120$\mu K \sqrt{s}$ @ 220 GHz, plus a 450 GHz dust monitor,
all with 12 arcmin FWHM beam).
The experiment will scan the Orologium region, the lowest dust
contrast region in the sky, which happens to be opposite
to the Sun in the Antarctic summer. 
The scan strategy and focal plane setup provide several confirmation
levels for the detected signals. At short times (few seconds), the
detected signal should reproduce in the second detector of the
same row, and both the detectors should see in time-reverse in the 
back-scan the structure seen in the forward-scan; a few minutes later
the second row of detectors should see the same structure.
All this should reproduce the day after, with only a slight
sky rotation, which becomes significant at the end of the flight.
The forecast for the 
final sensitivity of 15 $\mu K$ per 
20 arcminutes pixel at 90 GHz and 25 $\mu K$ per 12 arcminutes
pixel at 150 GHz, with a sky coverage of
2500 square degrees.

MAXIMA is developed by the University of California at Berkeley, with
contributions from Caltech, QMW, Universita' di Roma La Sapienza,
IROE-CNR Firenze. A 1.5 m off axis primary feeds an array 
receiver with 14 bolometers (8 at 150 GHz, 3 at 240 GHz,
3 at 420 GHz). All the detectors have a 11 arcmin FWHM.
Sky chop is obtained by nutating the lightweight carbon-fiber primary
mirror, thus obtaining a 4 deg/s, 6$^o$ amplitude scan of the beam.
In addition, a slower azimuth slew of the gondola (50$^o$ p-p) 
will produce a significant sky coverage. Two partially overlapping regions
will be scanned in a regular 6 hours north-America flight, in order 
to check for systematics and make easier the map reconstruction.
The experiment is expected to map about 26000 independent pixels,
with final sensitivity of $\sim 24 \mu K$/pixel at 150 GHz
and $\sim 66 \mu K$/pixel at 240 GHz. The flight of
MAXIMA will be in spring 1998.

TOP-HAT is a new-generation payload mounted on-top of the balloon,
and designed for Long Duration Ballooning. The system is developed
at NASA-Goddard and Bartol, U. of Chicago, U. of W. at Madison. 
The peculiar location of the
payload features better ground shields and absence of signals 
reflected from the balloon, allowing observations near the zenith.
The disadvantages are a stricter weight limit, a complex ACS and
the difficult recovery of the payload after test flights.
The experiment has a low sidelobes Cassegrain telescope with
a single pixel receiver (5 bands from 150 to 630 GHz, 20 arcmin FWHM beam). 
The test flight of TopHat Pointer (planned for the spring of 1998)
will cover the same NCP region observed by MSAM and Saskatoon,
with improved sensitivity and high frequency coverage.
In the LDB flight (1999), the telescope, tipped 12 degrees off the zenith, spins 
observing a 24 deg circle. Sky rotation allows the system to map
1800 square degrees of sky and about 20000 independent pixels.

\section{Conclusions}

Ballooning for CMB Anisotropy measurements is a very active field worlwide.
The activity is growing, for two main scientific reasons:

1) High frequency ($>$90 GHz) and high angular resolution ($\sim$ 10 arcmin FWHM) 
measurements are possible and effective.

2) These measurements complement the forthcoming data from MAP
and are a very important test-bed for Planck technologies.

Moreover, good science is being produced, and promises for
important results (like the $\ell$-space spectroscopy of
the acoustic peaks, or detection/falsification of non-gaussian statistics
for the CMB fluctuations) are quite convincing.

As an example relevant for this conference, we can mention the 
fact that determination of several cosmological parameters is
possible with very good accuracy from LDB experiments. For example,
if all the systematics effects are properly removed,
a single LDB experiment with 12 arcmin FWHM beams, 
16 total power bolometric detectors with sensitivity 
of 80 $\mu K \sqrt{s}$, 10 days of observing time
spent over a $50^o \times 50^o$ sky region, can measure the power
spectrum of the CMB anisotropies with very good accouracy \cite{Silvia}. 
Fits can be done on the measured power spectrum \cite{KL}, allowing to
recover $\Omega_{tot}$ with a 3$\%$ error, $\Omega_{\Lambda}$
with 6$\%$ error, $\Omega_B$ with $1\%$ error, $n$ scalar with
18$\%$ error, $H_o$ with 1$\%$ error, $Q_{rms,PS}$ with 4$\%$ error.
Here the errors for any parameter make no assumptions about the value
of the other parameters. These measurement errors can be significantly 
reduced if one or more of the cosmological parameters are constrained 
by other observations or fixed by assumptions.

\section*{Acknowledgments}

This work has been supported by Programma Nazionale Ricerche in Antartide,
Agenzia Spaziale Italiana, Universita' di Roma "La Sapienza". We thank
M. Devlin, S. Hanany, P. Meinhold, B. Netterfield, J. Ruhl 
for providing us with informations about several ongoing and planned
experiments.

\end{document}